\documentclass{article}

\usepackage{arxiv}

\usepackage[utf8]{inputenc} 
\usepackage[T1]{fontenc}    
\usepackage{hyperref}       
\usepackage{url}            
\usepackage{booktabs}       
\usepackage{amsfonts}       
\usepackage{nicefrac}       
\usepackage{microtype}      
\usepackage{lipsum}
\usepackage{graphicx}
\usepackage[square, numbers]{natbib}
\usepackage{afterpage}
\usepackage[table,xcdraw]{xcolor}
\usepackage{booktabs}
\usepackage{caption} 
\usepackage{enumitem}

\title{Characterizing Requirements Smells}

\author{
Emanuele Gentili\\
 MBDA Italy Spa, Rome, Italy \\
\texttt{emanuele.gentili@mbda.it}\\
\And
 Davide Falessi \\
 University of Rome Tor Vergata, Rome, Italy\\
  \texttt{falessi@ing.uniroma2.it} \\
}

\begin{document}
\maketitle
\begin{abstract}
\textbf{Context}: Software specifications are usually written in natural language and may suffer from imprecision, ambiguity, and other quality issues, called thereafter, requirement smells. Requirement smells can hinder the development of a project in many aspects, such as delays, reworks, and low customer satisfaction. From an industrial perspective, we want to focus our time and effort on identifying and preventing the requirement smells that are of high interest.  
\textbf{Aim}: This paper aims  to characterise 12 requirements smells in terms of frequency, severity, and effects.
\textbf{Method}: We interviewed ten experienced practitioners from different divisions of a large international company in the safety-critical domain called MBDA Italy Spa. 
\textbf{Results}: Our interview shows that the smell types perceived as most severe are Ambiguity and Verifiability, while as most frequent are Ambiguity and Complexity. We also provide a set of six lessons learnt about requirements smells, such as that effects of smells are expected to differ across smell types.
\textbf{Conclusions}: Our results help to increase awareness about the importance of requirement smells. Our results pave the way for future empirical investigations, ranging from a survey confirming our findings to controlled experiments measuring the effect size of specific requirement smells. 
\end{abstract}


\section{Introduction}
\label{sec:introduction}
Software requirements specifications are usually written in natural language  \citep{DBLP:books/daglib/0024869, kassab2014state} and may suffer from imprecision, ambiguity, and other quality issues, called thereafter, requirement smells \citep{mavin2010big}. Requirement smells can hinder the development of a project in many aspects, such as delays, reworks, and low customer satisfaction \citep{ahonen2010software, DBLP:books/daglib/0025717}.  

Researchers identified many types of smells and developed mechanisms, such as tools or regular expressions, for smell identification \citep{DBLP:journals/corr/abs-2305-07097, DBLP:conf/se/MontgomeryFBSM23}. However, removing all smells is expensive given the high impact of the change on many artefacts such as design, code, testing, or certification \citep{DBLP:journals/jss/KretsouAADG21}. Knowing which smell is important for whom, when, and why reasonably supports the reduction and prevention of smells. Therefore, it is key to gain insights into different types of smells.  

\citet{DBLP:journals/ese/FernandezWKFMVC17} reported a survey on the current status and issues in the requirements engineering process. We share with \citet{DBLP:journals/ese/FernandezWKFMVC17} the need to gather additional empirical evidence about requirements and their quality. However, while they focus on the requirements engineering process, we focus on the requirements as artefacts written in natural language.

\citet{DBLP:conf/se/MontgomeryFBSM23} reported a comprehensive mapping study on defining, improving, or evaluating requirements quality. However, to the best of our knowledge, no study investigated if and how the frequency, severity, or effects change across types of smells. \citet{DBLP:conf/se/MontgomeryFBSM23} acts as our baseline to identify 12 types of smells: Ambiguity, Completeness, Consistency, Correctness, Complexity, Traceability, Reusability, Understandability, Redundancy, Verifiability, Relevancy, and Undefined.  

We share the view of \citet{DBLP:journals/corr/FemmerF0E16} that "Whether a Requirements Smell finding is or is not a problem in a certain context must be individually decided for that context and is subject to reviews and other follow-up quality assurance activities." Thus, this research stems from the industrial need to focus our time and effort on removing and preventing the specific requirement smells that are of high interest. 

The aim of this paper is to characterise 12 requirement smells in terms of frequency, severity, and effects. To the best of our knowledge, no previous study analysed how frequency, severity, or effects vary across requirement smells.

We interviewed ten experienced practitioners from different divisions of a large international company in the safety-critical domain called MBDA Italy Spa.  

Our interview shows that the smell types perceived as most severe are Ambiguity and Verifiability while as most frequent are Ambiguity and Complexity. We also provide a set of six lessons learned about requirement smells such as that effects of smells are expected to differ across smell types.  

The remainder of this paper is structured as follows. Section \ref{sec:related} discusses the related literature, focusing in particular on requirements and their smells. Section \ref{sec:design} reports the design, Section \ref{sec:results} the results. Finally, Section \ref{sec:conclusions} concludes the paper and outlines directions for future work. 
\section{Related Work}
\label{sec:related}
A "requirement" refers to a specific functionality, constraint or quality that a system must possess in order to meet the needs of its users and stakeholders\citep{DBLP:books/daglib/0024869}. Since stakeholders' points of view may differ significantly from each other, and a common language is needed in order to communicate and share information among parties, requirements are usually expressed in natural language \citep{DBLP:books/daglib/0024869,sommerville1997requirements, robertson2012mastering}. 

\citet{kassab2014state} report that 61\% of users prefer to express requirements in natural language, whereas only 33\% use other semi-formal notations like UML. Despite the acquaintance a user can have with natural language, a non-systematic approach, i.e. unstructured, is likely to induce smells on the requirements specification, such as ambiguity, incompleteness, inconsistency and incorrectness \citep{DBLP:journals/ese/FernandezWKFMVC17}. These smells reasonably cause problems during the process stages and, lately, can determine the success or the failure of a project\citep{ahonen2010software, DBLP:books/sp/DickHJ17}. 

\citet{DBLP:conf/seke/SubramaniamLFE04} and \citet{mencl2004deriving} propose approaches to prevent undesired effects of smells by means of limiting, i.e. structuring, the natural language syntax, to the extent of automatically generating use cases models for Object Oriented languages. Similarly, \citet{DBLP:journals/corr/FemmerF0E16} and \citet{DBLP:journals/ese/FerrariGRTBFG18} propose tools to automatically identify smells in requirements descriptions according to software requirements definition norms (like CENELEC EN 50128:2011) \citep{DBLP:journals/ese/FerrariGRTBFG18} or standards (like INCOSE or ISO 29184) \citep{DBLP:journals/corr/FemmerF0E16}. These tools aim at driving the requirement elicitation process and assuring higher confidence in the requirement's quality. 

In order to gain a deeper insight into Requirement Engineering state-of-the-art, \citet{DBLP:conf/se/MontgomeryFBSM23} conducted a systematic mapping study on 105 relevant primary studies that use "empirical research to define, improve, or evaluate quality attributes". They identified 12 quality attribute themes, specified in 111 attributes sub-types, and reported that most of the studies concentrated on ambiguity, completeness, consistency and correctness quality attribute themes (63\%). We share their quality attributes categorization and used them as categories for requirement smells.  

\subsection{Automated requirement smells detection}
Requirement Engineering is acknowledged as an expensive, time-consuming, and error-prone process \citep{DBLP:journals/tse/Boehm84, DBLP:books/daglib/0067101}. This is especially true for complex systems with numerous requirements, e.g. thousand of requirements, making it challenging to obtain a comprehensive project specification overview. As a solution, automating smell detection becomes necessary in this context. 

As reported by \citet{DBLP:conf/se/MontgomeryFBSM23}, a total of 41 distinct tools have been developed to detect requirement smells, and aspects such as ambiguity, incompleteness and inconsistency resulted as the most studied. For instance, \citet{DBLP:journals/corr/FemmerF0E16} focused on analyzing the syntax of requirements expressed in natural language, providing a tool, called \textit{Smella}, implementing part of the Requirement Engineering standard ISO, IEC, and IEEE. ISO/IEC/IEEE 29148:2011 \citep{iso2011ieee}, and assessed the usefulness of the tool in the requirement elicitation process. Similarly, \citet{DBLP:conf/re/SekiHS19} designed a tool for detecting 22 "bad smells" (i.e. requirement smell sub-attributes) in use case description using structured natural language, achieving good performance in terms of precision and recall. 

\citet{DBLP:journals/corr/abs-2305-07097} developed a tool called \textit{Paska} based on Rimay\citep{DBLP:journals/ese/VeizagaATSB21}, and conducted an industrial case study on 13 system requirements specification documents from information system in financial domain, achieving a precision and recall in detecting smells of 89\%. All the authors agree on the importance of assessing the usefulness of the developed tools through direct feedback from practitioners and on the necessity of a larger empirical evaluation.


\subsection{Empirical evaluation}

To the best of our understanding, the closest empirical evaluation to our study is \citep{DBLP:journals/ese/FernandezWKFMVC17}. If, on the one side, we share their need to gather more empirical data about requirements quality,  on the other side, we differ in many aspects, such as the approach (survey vs interview) and the object under evaluation. Specifically, they focus on the requirements engineering process, whereas we focus on the requirements artefact as written in natural language. This allows us a way to create a preliminary knowledge base of which smells we should focus on the most. 

Regarding the effectiveness of elicitation techniques, \citet{DBLP:conf/re/DavisTHJM06} conducted a systematic review, reporting that \textit{interview} is the most commonly used elicitation technique, albeit there are no studies assessing that it is the most effective choice. Moreover, across interview strategies, the structured interview is the one gathering more information than unstructured interviews, sorting and ranking or thinking aloud techniques. 

This turns out to be even more important if we consider the study conducted by  \citep{DBLP:conf/fgit/SolemonSG09}, in which it emerges that companies with a "high-maturity rating", i.e. companies claiming to follow the best Requirement Engineering practices as a part of their quality management process, experience the same Requirement Engineering problems of companies with lower scores, remarking the necessity of looking deeply inside Requirement Engineering practices.


\section{Methodology}
\label{sec:design}
In this section we report on the methodology we use in this work.

\subsection{Industrial context}

MBDA Spa is a multinational defence company specialising in the defensive and aerospace domain. We work closely with armed forces and defence organizations to provide advanced defence solutions. Our expertise lies in research, development, and integration of cutting-edge technologies to enhance national security and contribute to the defence capabilities of their client nations. 
The company comprises four national companies located in Italy, France, Germany, and United Kingdom. To conduct this study, we interviewed ten individuals from MBDA Italy Spa, which serves as the central site for software development supporting all the company's solutions.

\afterpage{
\begin{table}[]
\centering
\caption{Population and project characterization.}
\resizebox{0.72\textwidth}{!}{

\begin{tabular}{|lll|llllll|}
\hline
\rowcolor[HTML]{808080} 
\multicolumn{3}{|c|}{\cellcolor[HTML]{808080}{\color[HTML]{FFFFFF} \textbf{Interviewees characteristics}}}                                                                                                                   & \multicolumn{6}{c|}{\cellcolor[HTML]{808080}{\color[HTML]{FFFFFF} \textbf{Project  characteristics}}}                                                                                                                                                                                                                                                                                                                                                                                          \\ \hline
\rowcolor[HTML]{A5A5A5} 
\multicolumn{1}{|l|}{\cellcolor[HTML]{A5A5A5}{\color[HTML]{FFFFFF} \textbf{Id}}} & \multicolumn{1}{l|}{\cellcolor[HTML]{A5A5A5}{\color[HTML]{FFFFFF} \textbf{Role}}} & {\color[HTML]{FFFFFF} \textbf{\#YE}} & \multicolumn{1}{l|}{\cellcolor[HTML]{A5A5A5}{\color[HTML]{FFFFFF} \textbf{\#Req}}} & \multicolumn{1}{l|}{\cellcolor[HTML]{A5A5A5}{\color[HTML]{FFFFFF} \textbf{\#Dev}}} & \multicolumn{1}{l|}{\cellcolor[HTML]{A5A5A5}{\color[HTML]{FFFFFF} \textbf{\#LOC}}} & \multicolumn{1}{l|}{\cellcolor[HTML]{A5A5A5}{\color[HTML]{FFFFFF} \textbf{\#YP}}} & \multicolumn{1}{l|}{\cellcolor[HTML]{A5A5A5}{\color[HTML]{FFFFFF} \textbf{\#Exc}}} & {\color[HTML]{FFFFFF} \textbf{Domain}} \\ \hline
\multicolumn{1}{|l|}{I1}                                                         & \multicolumn{1}{l|}{SGL}                                                                  & 25                                             & \multicolumn{1}{l|}{1000}                                                                    & \multicolumn{1}{l|}{4}                                                              & \multicolumn{1}{l|}{300K}                                                           & \multicolumn{1}{l|}{3}                                                                & \multicolumn{1}{l|}{1}                                                              & SRT                                    \\ \hline
\multicolumn{1}{|l|}{I2}                                                         & \multicolumn{1}{l|}{SGL}                                                                  & 22                                             & \multicolumn{1}{l|}{400}                                                                     & \multicolumn{1}{l|}{10}                                                             & \multicolumn{1}{l|}{100K}                                                           & \multicolumn{1}{l|}{2}                                                                & \multicolumn{1}{l|}{2}                                                              & SRT                                    \\ \hline
\multicolumn{1}{|l|}{I3}                                                         & \multicolumn{1}{l|}{SPL}                                                                  & 7                                              & \multicolumn{1}{l|}{2000}                                                                    & \multicolumn{1}{l|}{12}                                                             & \multicolumn{1}{l|}{250K}                                                           & \multicolumn{1}{l|}{6}                                                                & \multicolumn{1}{l|}{18}                                                             & SRT                                    \\ \hline
\multicolumn{1}{|l|}{I4}                                                         & \multicolumn{1}{l|}{Tx}                                                                   & 21                                             & \multicolumn{1}{l|}{300}                                                                     & \multicolumn{1}{l|}{7}                                                              & \multicolumn{1}{l|}{250K}                                                           & \multicolumn{1}{l|}{2}                                                                & \multicolumn{1}{l|}{15}                                                             & SRT                                    \\ \hline
\multicolumn{1}{|l|}{I5}                                                         & \multicolumn{1}{l|}{SGL}                                                                  & 21                                             & \multicolumn{1}{l|}{750}                                                                     & \multicolumn{1}{l|}{7}                                                              & \multicolumn{1}{l|}{70K}                                                            & \multicolumn{1}{l|}{3}                                                                & \multicolumn{1}{l|}{10}                                                             & SRT                                    \\ \hline
\multicolumn{1}{|l|}{I6}                                                         & \multicolumn{1}{l|}{SWEng}                                                                & 3                                              & \multicolumn{1}{l|}{200}                                                                     & \multicolumn{1}{l|}{3}                                                              & \multicolumn{1}{l|}{8K}                                                             & \multicolumn{1}{l|}{5}                                                                & \multicolumn{1}{l|}{1}                                                              & HRT                                    \\ \hline
\multicolumn{1}{|l|}{I7}                                                         & \multicolumn{1}{l|}{HoD/Tx}                                                               & 23                                             & \multicolumn{1}{l|}{400}                                                                     & \multicolumn{1}{l|}{12}                                                             & \multicolumn{1}{l|}{50K}                                                            & \multicolumn{1}{l|}{4}                                                                & \multicolumn{1}{l|}{20}                                                             & SRT                                    \\ \hline
\multicolumn{1}{|l|}{I8}                                                         & \multicolumn{1}{l|}{SWEng}                                                                & 18                                             & \multicolumn{1}{l|}{2000}                                                                    & \multicolumn{1}{l|}{12}                                                             & \multicolumn{1}{l|}{250K}                                                           & \multicolumn{1}{l|}{6}                                                                & \multicolumn{1}{l|}{18}                                                             & SRT                                    \\ \hline
\multicolumn{1}{|l|}{I9}                                                         & \multicolumn{1}{l|}{SGL}                                                                  & 12                                             & \multicolumn{1}{l|}{100}                                                                     & \multicolumn{1}{l|}{6}                                                              & \multicolumn{1}{l|}{70k}                                                            & \multicolumn{1}{l|}{5}                                                                & \multicolumn{1}{l|}{1}                                                              & HRT                                    \\ \hline
\multicolumn{1}{|l|}{I10}                                                        & \multicolumn{1}{l|}{SWEng}                                                                & 16                                             & \multicolumn{1}{l|}{2000}                                                                    & \multicolumn{1}{l|}{12}                                                             & \multicolumn{1}{l|}{250k}                                                           & \multicolumn{1}{l|}{6}                                                                & \multicolumn{1}{l|}{18}                                                             & SRT                                    \\ \hline
\end{tabular}}
\label{table:population_and_project_characterization}
\end{table}
}

\subsection{Study design}
In this work we use a qualitative semi-structured interview method, which has been proven to be a flexible instrument for investigating areas of interest whose boundaries are not clear nor complete \citep{robson2002real}.
Knowing how much a smell is frequent and severe reasonably supports reducing and preventing smells. To investigate which smell is particularly severe or frequent, we used the approach adopted by \citet{DBLP:journals/ese/FernandezWKFMVC17}. Specifically, we asked about the three top and least severe smells and about the three top and least frequent smells.  

Concerning types of smells, we use the categorization proposed by \citet{DBLP:conf/se/MontgomeryFBSM23}: \textit{Ambiguity}, \textit{Completeness}, \textit{Complexity}, \textit{Consistency}, \textit{Correctness}, \textit{Traceability}, \textit{Reusability}, \textit{Understandability}, \textit{Redundancy}, \textit{Verifiability}, \textit{Relevancy} and \textit{Undefined}. We refer to \citet{DBLP:conf/se/MontgomeryFBSM23} for their definitions. \\

\noindent Results about population and project characterization are reported in Table \ref{table:population_and_project_characterization}.

\noindent The list of questions is hereafter reported:
\begin{itemize}
    \item \textbf{Interviewee characterization}:
    \begin{enumerate}
        \item What is your current role? (Role)
        \begin{itemize}
            \item SW Engineer (SWEng): designs, develops, and maintains software systems for various applications.
            \item Technical Expert (Tx): provides specialized knowledge and expertise in software requirements management and architecture modelling. 
            \item SW Project Leader (SPL): leads a software project, coordinates teams, manages resources, and ensures successful delivery of high-quality software solutions.
            \item SW Group Leader (SGL): leads and coordinates a group of projects within a specific field, facilitating Software Project Leaders to ensure successful project execution and delivery.
            \item Head of Department (HoD): leads the software department, setting strategic direction, manages SW Group Leaders and Project Leaders, and ensures efficient software development operations.
        \end{itemize}
        \item How many years of experience do you have in software development?(\#YE)
    \end{enumerate}
    \item \textbf{Project characterization}: we asked information regarding projects that the interviewees are currently engaged in, or, if working on more than one, for a project they perceive as noteworthy in terms of requirement smells analysis.
    \begin{enumerate}
        \item How many requirements does the project consist of?(\#Req)
        \item How many developers does the team consist of?(\#Dev)
        \item To which domain does the project belong to?(Domain)
        \begin{itemize}
            \item Soft Real Time (SRT): refers to systems where meeting timing constraints is important but not critical. Occasional delays or missed deadlines may be tolerable as long as the overall system performance remains acceptable.
            \item Hard Real Time (HRT): refers to systems where meeting strict timing constraints is crucial, and failure to do so can result in catastrophic consequences.
        \end{itemize}
        \item How many software components, as the number of executables, does the project consist of?(\#Exc)
        \item How many Line Of Code does the project consist of?(\#LOC)
        \item How many years does the project last?(\#YP)
    \end{enumerate}

    \item \textbf{Requirement smells}:
    \begin{enumerate}
        \item What are the three most and least severe requirement smells? 
        \item What are the three most and least frequent requirement smells? 
        \item What are the effects of a certain requirement smell?  
        \item Are there contexts in which the effects of a certain smell result mitigated/amplified? 
    \end{enumerate}
\end{itemize}

\subsection{Validity}
Since the results rely on a small set of interviews from a single company, we recommend care in generalising results in other contexts. 

There might be threats to validity even within our company. For instance, our population might not be representative of the company. To mitigate this threat, we selected subjects with a representative proportion of roles to face this threat.

Another possible threat to validity is selection bias, i.e., that the selected subjects might be biased towards specific answers. We believe this threat is negligible since we had a 100\% acceptance rate and subjects with a representative proportion of roles. 

An additional possible threat to validity is a wrong interpretation of subjects' answers. To face this threat, we provide no pressure on the time or direction of the answers. We also adopted a semi-structured interview protocol, which allowed us to spot possible misunderstandings while diving deep towards an answer. We also analysed the subjects' answers multiple times to ensure nothing was forgotten or misinterpreted.

\section{Study Results}
\label{sec:results}
In the following we report the lessons learnt as extracted by analysing our interviews.

\subsection{LL1: The perceived severity varies across types of smells.}
Figure \ref{fig:smellseveritydistribution} reports the frequency distribution of the three most and least severe requirement smells. According to Figure \ref{fig:smellseveritydistribution},  the perceived most severe smells are Ambiguity (80\%), Verifiability (80\%), and Consistency (60\%). The perceived least severe smells are Relevancy (90\%), Reusability (70\%) and Redundancy (50\%). Interestingly, only Completeness and Correctness have been identified as the most and least three severe by different interviewees. 

\afterpage{%
  \begin{figure}[t!] 
    \centering
        \includegraphics[width=0.7\textwidth]{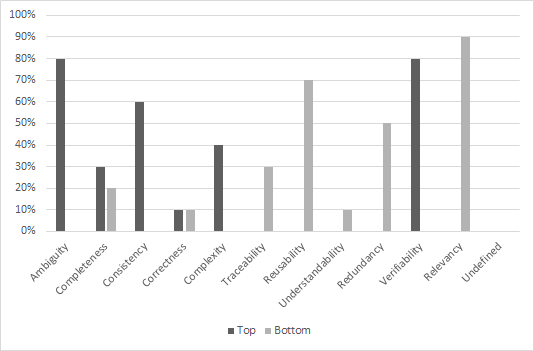}
        \rule{32.9em}{0.3pt}
    \caption{Distribution of the three most and least severe requirement smells.}
    \label{fig:smellseveritydistribution} 
  \end{figure}
} 

\subsection{LL2: The perceived severity of the same smell varies across project domains}
We know that requirement smells can cause rework and time and cost overruns \citep{ahonen2010software, DBLP:books/daglib/0025717}, and in some cases, a smell might even be catastrophic. For instance, requirements concerning the performance of the system are key for HRT systems \citep{laplante2004real}. A single smell in a performance requirement of an HRT system might have a huge impact on the overall project or even people's lives. For instance, regarding Verifiability, "If a performance requirement does not come with clear time constraints, we have a huge verifiability problem. For instance, if a computation task is described with high priority and to be executed fast, this description does not lead to clear tests and therefore, the system might pass the test and eventually create system malfunction since the actual priority and speed constraints required by the production context differ from the tested ones." (cit. I9)

\subsection{LL3: The perceived frequency varies across types of smells.}
Figure \ref{fig:smellfrequencydistribution} reports the frequency distribution of the most and least frequent requirement smells. According to Figure \ref{fig:smellfrequencydistribution}, the perceived most frequent smells are Ambiguity (70\%), Complexity (70\%) and Consistency (40\%). The perceived least frequent smells are Understandability (60\%), Reusability (40\%), and Relevancy (50\%). We note that differently from severity, the majority of smells are perceived as most and least frequent by different interviewees; this suggests less agreement among interviewees or, likely, the presence of other factors influencing the frequency of smells, such as roles or phases. 

\afterpage{%
  \begin{figure}[t!] 
    \centering
        \includegraphics[width=0.7\textwidth]{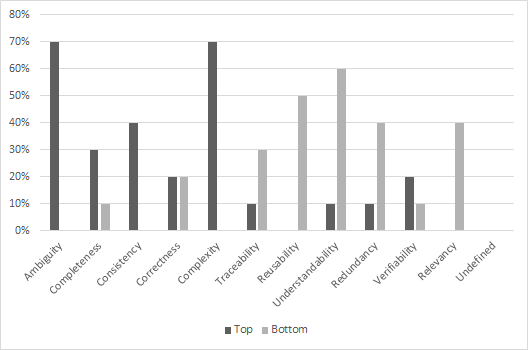}
        \rule{32.9em}{0.3pt}
    \caption{Distribution of the top three and bottom three requirement smell frequency.}
    \label{fig:smellfrequencydistribution} 
  \end{figure}
} 

\subsection{LL4: The frequency of a smell is perceived differently across roles or phases.}
The requirements are managed over the development life cycles and get improved over the life cycle. Different companies have a proportion of different roles; each role gets the requirement at a different stage and hence at a different quality. Thus, some smells might not be frequent for a role because another role has already fixed the smell. For instance, “The presence of a SW Requirement Specification expert, who centralizes and pre-filters requirements affected by smells, is fundamental for reducing the frequency of such smells during the Development phase, and helps the whole Team to stay aligned with the given specification.” (cit. I8)

\subsection{LL5: The perceived effects may vary across types of smells.}
We know that requirements smell can cause rework and time and cost overruns \citep{ahonen2010software, DBLP:books/daglib/0025717}. Moreover, reasonably, specific smells cause specific problems. However, the specific effects of specific requirement smells, to the best of our knowledge, are unknown.  Let's discuss the examples of the Verifiability, Ambiguity, and Complexity smells.

Regarding Verifiability, if it is unclear how to verify a requirement, then the verification might be incorrect and hence will likely require to be performed multiple times, thus impacting the time and cost of testing. An additional effect of the Verifiability smell is that bugs might not be found during testing, thus leading to decreased customer satisfaction and increased development costs. For instance,  "The Verifiability of requirements can determine the success or the failure of a project: scarcely verifiable requirements determine special effort during the Coding phase (it is not clear how to code in order to provide evidence of the desired behaviour) and during the Testing phase (in fact the number and complexity of Test Cases grow significantly)... and a poorly tested SW is likely to exhibit bugs during the Maintenance phase, leading to a high impact in rework, time, extra costs and customer satisfaction, with a general loss of credibility of the Company." (cit. I2) 

Regarding Ambiguity, if a requirement is unclear, this will likely need to go back and forth between requirements engineers and developers to identify and formalise a clear version of the content. Thus, an ambiguous requirement is the subject of many change requests. Specifically, "Across all the smells, Ambiguity is the one causing more problems: if evident, it can be addressed and solved at an early stage, before starting to develop code, with relatively little impact in terms of rework; but sometimes, it remains uncaught until Integration Test stage (or even worst, until Maintenance stage) causing bugs whose resolution will have, possibly, a very high impact in term of rework on all process stages, on costs and customer satisfaction." (cit. I2) 

Finally, regarding Complexity, it might be that a complex description of functionality leads to a complex implementation of that functionality. Specifically, "When a requirement is too Complex, one of the main effects is that practitioners tend to implement code with the same degree of complexity."  (cit. I7) 

We note that Verifiability and Ambiguity do not reasonably lead to complex code. Similarly, Complexity does not reasonably lead to decreased customer satisfaction. Thus, the impact of requirement smells is perceived as varying across smells.

\subsection{LL6: The severity of a smell might change across the stage of the project.} 
We know that requirement smells can have negative effects on software development \citep{ahonen2010software, DBLP:books/daglib/0025717}; however, we realize that the effects of smells might be null or even positive in some circumstances \citep{DBLP:journals/corr/FemmerF0E16}. Let’s have an example of how the effect of the underspecification\citep{DBLP:conf/se/MontgomeryFBSM23}, a sub-smell of the Completeness smell, changes across the development process stages and can even be positive. Regarding underspecification, we know from the literature that requirements get more specified over time as the clients get a better understanding of what they want \citep{DBLP:journals/jss/Kruchten08}. Thus, some needs that are underspecified at the early stage of the project get specified over time; other needs might remain underspecified since the clients do not (need to) provide more details. Thus, the roles approaching an underspecified requirement at the early stage of the development process are in trouble since they need to make decisions based on assumptions that might change when clients will better specify their needs \citep{DBLP:journals/csur/FalessiCKK11}. Counter-wise, roles approaching an underspecified requirement at a late stage are glad to have many options, knowing that no additional details will invalidate the chosen solution. Specifically, on the one side, “An underspecified functional requirements at an early stage can cause the SW architect to design an incorrect architecture with little-flexibly, not able to satisfy constraints that will come up later during the development stage. So it can potentially bring to the redesign of the whole architecture, at the price of losing time, money and increasing the frustration of the whole development team.” (cit. I3)  On the other side, “A too-abstract requirement is not necessarily bad news: from Project Leader and Technical Expert points of view, it provides a high degree of freedom in terms of selection of the most convenient software architecture, with the possibility to experiment with newest, and more adequate, SW solutions.” (cit. I4)  We do not know if this reasoning applies to requirements sub-smells other than underspecification.  

\section{Conclusions}
\label{sec:conclusions}
From an industrial perspective, we want to focus our time and effort on identifying and preventing the requirement smells that are important. Knowing which smell is important for whom, when, and why reasonably supports the reduction and prevention of smells.

Our results rely on ten industrial experts and reveal that the smell types perceived as most important are Ambiguity and Verifiability, while as most frequent are Ambiguity and Complexity. We also provide a set of six lessons learned about requirement smells, such as that effects of smells are expected to differ across types. Our results help increase awareness about the importance of requirement smells. 

To our best understanding, this study is the first attempt to characterise requirement smells in terms of severity, frequency and effects. Since the results rely on a small set of interviews from a single company, we recommend care in generalising results in other contexts. 

We note that correlation does not imply causation. Smells perceived as important or related to effects are probably only correlated to rather than causing effects. The concept of inherent complexity is something we know for code smells \cite{DBLP:journals/tse/SjobergYAMD13}  and likely applies to requirement smells too. Specifically, if something is easy to express in natural language, it is likely easy to design, code and test. Counter-wise, if something is complex, it remains complex regardless of how much time we spend improving its description. In other words, spending a lot of effort describing something complex might decrease the requirement smells, but it might not decrease the complexity of developing it. Of course, something easy might become complicated if described in a complex way. Thus, our results pave the way for future empirical investigations, ranging from a survey confirming our findings to mining software repositories to establish correlations between smells and effects on projects to controlled experiments measuring the size of smells effects.

\bibliographystyle{plainnat}  
\bibliography{references}

\end{document}